%% file: paper.tex
\documentclass[a4paper,10pt]{article}
\pdfoutput=1

\usepackage{acronym}
\usepackage[backend=bibtex]{biblatex}
\usepackage[usenames,dvipsnames]{color}
\usepackage{comment}
\usepackage{graphicx}
\usepackage{hyperref}
\usepackage[utf8]{inputenc}
\usepackage{listings}

\input{acronyms}
\bibliography{bibliography.bib}

\definecolor{Gray}{gray}{0.5}
\definecolor{OliveGreen}{cmyk}{0.64,0,0.95,0.40}

\title{Practical Concurrent Priority Queues}
\author{Jakob Gruber}

\begin{document}

\maketitle

\begin{abstract}
Priority queues are abstract data structures which store a set of key/value pairs
and allow efficient access to the item with the minimal (maximal) key. Such queues are an important
element in various areas of computer science such as algorithmics (i.e. Dijkstra's shortest
path algorithm) and operating system (i.e. priority schedulers).

The recent trend towards multiprocessor computing requires new implementations of basic
data structures which are able to be used concurrently and scale well to a large number
of threads. In particular, lock-free structures promise superior scalability by avoiding
the use of blocking synchronization primitives.

Concurrent priority queues have been extensively researched over the past decades.
In this paper, we discuss three major ideas within the field: fine-grained locking
employs multiple locks to avoid a single bottleneck within the queue; SkipLists
are search structures which use randomization and therefore do not require elaborate reorganization
schemes; and relaxed data structures trade semantic guarantees for improved scalability.
\end{abstract}

\section{Introduction}


In the past decade, advancements in computer performance have been made mostly
through an increasing number of processors instead of higher clock speeds.
This development necessitates new approaches to data structures and algorithms
that take advantage of concurrent execution on multiple threads and processors.


This paper focuses on the priority queue data structure, consisting of two operations
traditionally called \lstinline|Insert| and \lstinline|DeleteMin|. \lstinline|Insert|
places an item into the queue together with its priority, while \lstinline|DeleteMin|
removes and returns the lowest priority item. Both of these operations are expected to have
a complexity of at most $O(\log n)$. Priority queues are used in a large variety
of situations such as shortest path algorithms and scheduling. 


Concurrent priority queues have been the subject of research since the 1980s
\cite{ayani1990lr,biswas1987simultaneous,das1996distributed,deo1992parallel,huang1991evaluation,
luchetti1993some,mans1998portable,olariu1991optimal,prasad1995parallel}.
While early efforts have focused mostly on parallelizing Heap structures
\cite{hunt1996efficient},
more recently priority queues based on \citeauthor{pugh1990skip}'s SkipLists
\cite{pugh1990skip} seem to show more potential \cite{shavit2000skiplist,
sundell2003fast,herlihy2012art,linden2013skiplist}. Current research has also examined
relaxed data structures \cite{wimmer2013data,alistarhspraylist} which trade
strictness of provided guarantees against improved scalability.


In the following, we investigate the evolution of concurrent priority queues suitable for general use,
i.e. unbounded range queues based on widely available atomic primitives such as \ac{CAS}.
Section \ref{sec:concepts} outlines basic concepts and definitions. In Section \ref{sec:hunt},
we cover the \citeauthor{hunt1996efficient} queue as a representative of early heap-based queues using
fine-grained locking to avoid the bottleneck of a single global lock. Lock-free SkipList-based
structures offering better disjoint-access parallelism are discussed in Section \ref{sec:lockfree},
while Section \ref{sec:relaxed} introduces the concept of relaxed data structures and presents two
such priority queues. Finally, related work is presented in Section \ref{sec:related}, and the paper is
concluded in Section \ref{sec:conclusion}.

\section{Concepts and Definitions} \label{sec:concepts}


Concurrent data structures are intended to be accessed simultaneously by several processes
at once. \emph{Lock-based} structures ensure that only a limited number of processes may enter
a critical section at once. \emph{Lock-free} data structures eschew the use of locks, and guarantee
that at least a single process makes progress at all times. Since lock-free structures are
non-blocking, they are not susceptible to priority inversion, deadlock, and livelock.
\emph{Wait-freedom} further guarantees that every process finishes each operation in a bounded number of steps.
In practice, wait-freedom often introduces an unacceptable overhead; lock-freedom
however has proven to be both efficient and to scale well to large numbers of processes.
Recently, \citeauthor{kogan2012methodology} have also developed a methodology for implementing efficient
wait-free data structures \cite{kogan2012methodology}.


There are several different criteria which allow reasoning about the correctness of concurrent
data structures. \emph{Linearizable} \cite{herlihy1990linearizability} operations appear to take
effect at a single instant in time at so-called linearization points.
\emph{Quiescently consistent} \cite{shavit1996diffracting} data structures
guarantee that the result of a set of parallel operations is equal to the result of a sequential ordering
after a period of quiescence, i.e. an interval without active operations has passed.
Linearizability as well as quiescent consistency are composable ---
any data structure composed of linearizable (quiescently consistent) objects is also linearizable
(quiescently consistent).
\emph{Sequential consistency} \cite{lamport1979make} requires the result of a set of operations
executed in parallel to be equivalent to the result of some sequential ordering of the same
operations.


Lock-free algorithms and data structures are commonly constructed using synchronization primitives
such as \acf{CAS}, \ac{FAA}, \ac{FAO}, and \ac{TAS}. The \ac{CAS} instruction, which atomically
compares a memory location to an expected value and sets it to a new value if they are equal,
is implemented on most modern architectures and can be considered a basic building block of lock-free
programming. More exotic primitives such as \ac{DCAS} and \ac{DCSS} exist as well, but are not yet
widely available and require inefficient software emulations to be used.


An area in memory accessed frequently by a large number of processes is said to be \emph{contended}.
Contention is a limiting factor regarding scalability: concurrent reads and writes to the same
location must be serialized by the cache coherence protocol, and only a single concurrent \ac{CAS}
can succeed while all others must retry. \emph{Disjoint-access parallelism} is the concept of
spreading such accesses in order to reduce contention as much as possible.


Priority queues are an abstract data structure allowing insertion of items with a given priority
and removal of the lowest-priority item in logarithmic time. Search trees and Heaps (which are
flattened representations of complete trees such that each node's key is at least as large as
those of both children) are concrete data structures which are usually used to implement sequential
priority queues. However, both require fairly elaborate reorganization after
{\lstset{breakatwhitespace=true} \lstinline|DeleteMin|}
and/or \lstinline|Insert| operations, which are especially challenging to achieve in a concurrent
environment.


SkipLists \cite{pugh1990skip} have become increasingly popular in concurrent data structures because
they are both simple to implement, and exhibit excellent disjoint-access parallelism. In contrast to
search trees and Heaps, reorganization is not necessary since SkipLists rely on
randomization for an expected $O(\log n)$ complexity of \lstinline|Insert|, \lstinline|Search|
and \lstinline|Delete| operations.
A SkipList may be visualized as a set of linked lists with corresponding levels. The linked
list at level 0 is the sorted sequence of all objects in the SkipList,
and higher levels provide ``shortcuts''
into the SkipList such that a list of level $i + 1$ contains a subset of the objects in level
$i$. A SkipList node $n$ is said to be of level $i$ if it is in all lists of levels $[0, i]$
and in none of levels $[i + 1, \infty]$. Upon insertion, the new node's level is assigned at random
according to a geometric distribution; deletion simply removes the node.

\begin{figure}[ht]
\begin{lstlisting}
struct slist_t {
  size_t max_level;
  node_t head[max_level];
};

struct node_t {
  key_t key;
  value_t value;
  size_t level;
  node_t *next[level];
};

/* All operations are expected O(log n) time. */
void slist_insert(slist_t *l, key_t k, value_t v);
bool slist_delete(slist_t *l, key_t k, value_t &v);
bool slist_contains(slist_t *l, key_t k);
\end{lstlisting}
\caption{Basic SkipList structure and operations.}
\label{fig:basicsl}
\end{figure}

\section{Fine-grained Locking Heaps} \label{sec:hunt}

In the remaining paper, we will discuss implementations of several concurrent priority queue implementations.
Early designs have mostly been based on search trees \cite{boyar1994chromatic,johnson1991highly} and
heaps \cite{ayani1990lr,biswas1987simultaneous,das1996distributed,deo1992parallel,huang1991evaluation,
luchetti1993some,mans1998portable,olariu1991optimal,prasad1995parallel}.
We chose the priority queue by \citeauthor{hunt1996efficient} \cite{hunt1996efficient}
as a representative of early concurrent priority queues since it has been proven to
perform well \cite{shavit2000skiplist} in comparison to other efforts of the time such as \cite{nageshwara1988concurrent,ayani1990lr,yan1998lock}. 
It is based on a Heap
structure and attempts to minimize lock contention between threads by a) adding per-node
locks, b) spreading subsequent insertions through a bit-reversal technique, 
and c) letting insertions traverse bottom-up in order to minimize conflicts with
top-down deletions.

However, significant limitations to scalability remain. A global lock is required
to protect accesses to a variable storing the Heap's size which all operations
must obtain for a short time. Disjoint-access through bit-reversal breaks down
once a certain amount of traffic is reached, since only subsequent insertions
are guaranteed to take disjoint paths towards the root node. Note also that
the root node is a severe serial bottleneck, since it is potentially part of
every insertion path, and necessarily of every \lstinline|DeleteMin| operation.
Finally, in contrast to later dynamic SkipList-based designs, the capacity of Hunt Heaps
is fixed upon creation.

Benchmarking results in the literature have been mixed; a sequential priority
queue protected by a single global lock outperforms the \citeauthor{hunt1996efficient}
Heap in most cases \cite{hunt1996efficient,sundell2003fast}. Speed-up only occurs once
the size of the Heap reaches a certain threshold such that concurrency
can be properly exploited instead of being dominated by global locking overhead.

\section{Lock-free Priority Queues} \label{sec:lockfree}

Traditional data structures such as the Heap have fallen out of favor;
instead, SkipLists \cite{pugh1990skip,pugh1998concurrent} have become the focus
of modern concurrent priority queue research
\cite{shavit2000skiplist,sundell2003fast,herlihy2012art,linden2013skiplist,alistarhspraylist}.
SkipLists are both conceptual simple as well as simple to implement; they also exhibit
excellent disjoint-access parallelism properties, and do not require rebalancing due to their
reliance on randomization.

A state of the art lock-free SkipList implementation based on the \ac{CAS} instruction
by \citeauthor{fraser2004practical} \cite{fraser2004practical} is freely available\footnote{
\url{http://www.cl.cam.ac.uk/research/srg/netos/lock-free/}} under a BSD license.
\citeauthor{fraser2004practical} exploits unused pointer bits to mark nodes as logically
deleted, with physical deletion following as a second step.

SkipLists are dynamic data structures in the sense that they grow and shrink
at runtime. In consequence, careful handling of memory accesses and (de)allocations
are required. As an additional requirement, these memory management schemes must
themselves be both scalable and lock-free to avoid limiting the SkipList itself.
\citeauthor{fraser2004practical} in particular employs lock-free epoch-based garbage-collection,
which frees a memory segment only once all threads that could have seen a pointer to it have
exited the data structure.

\subsection{\citeauthor{shavit2000skiplist}} \label{sec:shavit}

\citeauthor{shavit2000skiplist} were the first to propose the use of SkipLists
for priority queues \cite{linden2013skiplist}. Their initial locking implementation
\cite{shavit2000skiplist} builds on \citeauthor{pugh1998concurrent}'s concurrent
SkipList \cite{pugh1998concurrent}, which uses one lock per node per level.

A crucial observation is that nodes which are only partially connected
do not affect correctness of the data structure. As soon as the first level (i.e. \lstinline|node.level[0]|)
has been successfully connected, a node is considered to be in the SkipList.
Therefore, both insertions and deletions can be split into steps --- insertions
proceed bottom-up while deletions proceed top-down. Locks are held only for the current level
which helps to reduce contention between threads.

\begin{figure}[ht]
\begin{lstlisting}
struct node_t {
  [...] /**< Standard node members as above. */
  atomic<bool> deleted; /**< Initially false. */
  time_t timestamp;
  lock_t locks[level + 1];
};
\end{lstlisting}
\caption{\citeauthor{shavit2000skiplist} structure.}
\label{fig:shavitsl}
\end{figure}

Likewise, deletions are split into a logical phase (atomically setting the \lstinline|node.deleted|
flag) and a physical phase which performs the actual pointer manipulations and can be seen as a simple
call to the underlying SkipList's \lstinline|sl_delete| function.

A \lstinline|DeleteMin| call starts the list head, and attempts to atomically set
the deletion flag using a \lstinline|CAS(node.deleted, false, true)| call (or equivalent constructs).
If it succeeds, the current node is physically deleted and returned to the caller. Otherwise,
\lstinline|node.next[0]| is set as the new current node and the procedure is repeated.
If the end of the list is reached, \lstinline|DeleteMin| returns false to indicate an empty list.


Note that so far this implementation is not linearizable: consider the case in which a
slow thread A is in the middle of a \lstinline|DeleteMin| call. Within this context, we refer to
the node with key $i$ as node $i$, or simply $i$. Several \lstinline|CAS|
operations have failed, and A is currently at node $j$.
A fast thread B then first inserts a node $i$, followed by a node $k$ such that
$i < j < k$, i.e. the former and latter nodes are inserted, respectively, before and after
thread A's current node. Assuming further that all nodes between $j$ and $k$ have already
been deleted, then thread A will return node $k$. This execution is not linearizable; however,
it is quiescently consistent since operations can be reordered to correspond to some sequential
execution at quiescent periods.

\citeauthor{shavit2000skiplist} counteract this by introducing a \lstinline|timestamp| for each
node which is set upon successful insertion. In this variant, each \lstinline|DeleteMin| operation
simply ignores all nodes it sees that have not been fully inserted at the time it was called.

Explicit memory management is required to avoid dereferencing pointers to freed memory areas
by other threads after physical deletion. This implementation uses a dedicated garbage collector
thread in combination with a timestamping mechanism which frees \lstinline|node|'s memory only
when all threads that might have seen a pointer to \lstinline|node| have exited the data structure.

\citeauthor{herlihy2012art} \cite{herlihy2012art} recently described and implemented a lock-free,
quiescently consistent version of this idea in Java. While mostly identical, notable differences are
that a) the new variant is based on a lock-free skiplist, b) the timestamping mechanism was not
employed and thus linearizability was lost, and c) explicit memory management is not required
because the Java virtual machine provides a garbage collector.

\subsection{\citeauthor{sundell2003fast}} \label{sec:sundell}

\citeauthor{sundell2003fast} proposed the first lock-free concurrent priority queue in
\citeyear{sundell2003fast} \cite{sundell2003fast}. The data structure is linearizable
and is implemented using commonly available atomic primitives \ac{CAS}, \ac{TAS}, and \ac{FAA}.
In contrast to other structures covered in this paper, this priority queue is restricted to
contain items with distinct priorities. Inserting a new item with a priority already contained
in the list simply performs an update of the associated value.
A real-time version is also provided which we will not discuss further (interested readers are
referred to \cite{sundell2003fast}).

\begin{figure}[ht]
\begin{lstlisting}
struct node_t {
  [...] /**< Standard node members as above. */
  size_t valid_level;
  node_t *prev;
};
\end{lstlisting}
\caption{\citeauthor{sundell2003fast} structure.}
\label{fig:sundellsl}
\end{figure}

The structure of each node is basically identical to Figure \ref{fig:basicsl}. However, \citeauthor{sundell2003fast}
exploit the fact that the two least significant bits of pointers on 32- and 64-bit systems
are unused and reuse these as deletion marks. A set least significant bit on a pointer signifies
that the current node is about to be deleted.
Reuse of \lstinline|node.level[i]| pointers
prevents situations in which a new node is inserted while its predecessor is being removed,
effectively deleting both nodes from the list. Likewise, the reuse of the \lstinline|node.value|
pointer ensures that updates of pointer values (which occur when a node with the inserted priority already exists)
handle concurrent node removals correctly.

As in the \citeauthor{shavit2000skiplist} priority queue, insertions proceed bottom-up while
deletions proceed top-down --- on the one hand, the choice of opposite directions reduces collisions
between concurrent insert and delete operations, while on the other hand removing nodes from top levels first
allows many other concurrent operations to simply skip these nodes, further improving performance.
\lstinline|node.valid_level| is updated during inserts to always equal the highest level of the SkipList
at which pointers in this node have already been set (as opposed to \lstinline|node.level|, which equals
the final level of the node).

A helping mechanism is employed whenever a node is encountered that has its deletion bit set, which attempts
to set the deletion bits on all next pointers and then removes the node from the current level. The
\lstinline|node.prev| pointer is used as a shortcut to the previous node, avoiding a complete retraversal
of the list.

This implementation uses the lock-free memory management invented by \citeauthor{valois1996lock}
\cite{valois1995lock,valois1996lock} and corrected by \citeauthor{michael1995correction}
\cite{michael1995correction}. It was chosen in particular because this scheme can guarantee validity
of \lstinline|prev| as well as all \lstinline|next| pointers. Additionally, it does not require a separate
garbage collector thread.

A rigorous linearizability proof is provided in the original paper \cite{sundell2003fast} which shows
linearization points for all possible outcomes of all operations.

Benchmarks performed by \citeauthor{sundell2003fast} show their queue performing noticeably better than both locking
queues from Sections \ref{sec:shavit} and \ref{sec:hunt}, and slightly better than a priority queue
consisting of a SkipList protected by a single global lock.

\subsection{\citeauthor{linden2013skiplist}} \label{sec:linden}

One of the most recent priority queue implementations was published by \citeauthor{linden2013skiplist}
in \citeyear{linden2013skiplist} \cite{linden2013skiplist}. They present a linearizable, lock-free concurrent priority
queue which achieves a speed-up of $30-80\%$ compared to other SkipList-based priority queues by
minimizing the number of \ac{CAS} operations within most \lstinline|DeleteMin| operations.

A priority queue implementation is called deterministic when the algorithm does not contain randomized elements.
It is called strict when \lstinline|DeleteMin| is guaranteed to return the minimal element currently within the queue
(in contrast to relaxed data structures which are discussed further in the next section).
All such priority queues share an inherent bottleneck, since all threads calling \lstinline|DeleteMin| compete
for the minimal element, causing both contention through concurrent \ac{CAS} operations on the same variable
as well as serialization effort by the cache coherence protocol for all other processor accessing the same cache
line.

\begin{figure}[ht]
\begin{lstlisting}
struct node_t {
  [...] /**< Standard node members as above. */
  atomic<bool> inserting;
};
\end{lstlisting}
\caption{\citeauthor{linden2013skiplist} structure.}
\label{fig:lindensl}
\end{figure}

In this implementation, most \lstinline|DeleteMin| operations only perform logical deletion by setting
the deletion flag with a single \ac{FAO} call; nodes are only deleted physically once a certain
threshold of logically deleted nodes is reached.

This mechanism requires a new invariant, in that the set of all logically deleted nodes must always
form a prefix of the SkipList. Recall that in the \citeauthor{sundell2003fast} queue, deletion flags
for node \lstinline|n| were packed into \lstinline|n.next| pointers, preventing insertion of new
nodes \emph{after} deleted nodes. This implementation instead places the deletion flag into the
lowest level \lstinline|next| pointer of the previous node, preventing insertions \emph{before}
logically deleted nodes.

Once the prefix of logically deleted nodes reaches a specified length (represented by \lstinline|BoundOffset|),
the first thread to observe this fact within \lstinline|DeleteMin| performs the actual physical
deletions by updating \lstinline|slist.head[0]| to point at the last logically deleted node with a
single \ac{FAO} operation. The remaining \lstinline|slist.head| pointers are then updated, and
all physically deleted nodes are marked for recycling.

Since at any time, the data structure contains a prefix of logically deleted nodes, all \lstinline|DeleteMin|
operations must traverse this sequence before reaching a non-deleted node. In general, reads of nonmodified
memory locations are very cheap; however, benchmarks in \cite{linden2013skiplist} have shown that
after a certain point, the effort spent in long read sequences significantly outweighs the reduced
number of \ac{CAS} calls. It is therefore crucial to choose \lstinline|BoundOffset| carefully, with the
authors recommending a prefix length bound of 128 for 32 threads.

The actual \lstinline|DeleteMin| and \lstinline|Insert| implementations are surprisingly simple.
Deletions simply traverse the list until the first node for which \lstinline|(ptr, d) = FAO((node.next[0], d), 1)|
returns a previously unset deletion flag (\lstinline|d = 0|) and then return \lstinline|ptr|.
Insertions occur bottom-up and follow the basic \citeauthor{fraser2004practical} algorithm \cite{fraser2004practical},
taking the separation of deletion flags and nodes into account. The \lstinline|node.inserting| flag
is set until the node has been fully inserted, and prevents moving the list head past a partially
inserted node. \citeauthor{fraser2004practical}'s epoch-based reclamation scheme \cite{fraser2004practical}
is used for memory management.

The authors provide high level correctness and linearizability proofs as well as a model for the
SPIN model checker. Performance has been shown to compare favorably to both
\citeauthor{sundell2003fast} and \citeauthor{shavit2000skiplist} queues, with throughput improved by
up to $80\%$.

\section{Relaxed Priority Queues} \label{sec:relaxed}

The body of work discussed in previous sections
creates the impression that the outer limits of strict, deterministic priority queues have been reached.
In particular, \citeauthor{linden2013skiplist} conclude that scalability is solely limited by \lstinline|DeleteMin|,
and that less than one modified memory location per thread and operation would have to be read
in order to achieve improved performance \cite{linden2013skiplist}.

Recently, relaxation of provided guarantees have been investigated as another method of reducing
contention and improving disjoint-access parallelism.
For instance, k-FIFO queues \cite{kirsch2012fast} have achieved considerable
speed-ups compared to strict FIFO queues by allowing {\lstset{breaklines,breakatwhitespace} \lstinline|Dequeue|} to return elements
up to $k$ positions out of order.

Relaxation has also been applied to concurrent priority queues with some success, and in the following
sections we discuss two such approaches.

\subsection{\citeauthor{wimmer2013data}} \label{sec:wimmer}

\citeauthor{wimmer2013data} presented the first relaxed, linearizable, and lock-free priority queue
in \cite{wimmer2013data}. It is integrated as a priority scheduler into their \emph{Pheet} task-scheduling
system, and an open-source implementation is available\footnote{\url{http://pheet.org}}.

Their paper presents several variations on the common theme of priority queues: a distributed work-stealing
queue which can give no guarantees as to global ordering since it consists of separate priority queues
at each thread; a relaxed centralized priority queue in which no more than $k$ items are missed
by any processor; and a relaxed hybrid data structure which combines both ideas and provides
a guarantee that no thread misses more than $kP$ items where $P$ is the number of participating threads.
In this section, we examine only the hybrid variant since it provides both the scalability of work-stealing
queues and the ordering guarantees of the centralized priority queue.

\begin{figure}[ht]
\begin{lstlisting}
struct globals_t {
  list_of_item_t global_list;
};

/* Thread-local items. */
struct locals_t {
  list_of_item_t local_list;
  pq_t prio_queue;
  size_t remaining_k;
};
\end{lstlisting}
\caption{\citeauthor{wimmer2013data} structure.}
\label{fig:wimmerq}
\end{figure}

The hybrid queue consists of a list of globally visible items and one local item list as well as a local
sequential priority queue per thread. The thread-local counter \lstinline|remaining_k| tracks how many more
items may be added to the local queue until all local items must be made globally visible to avoid
breaking guarantees.

Whenever an item is added, it is first added locally to both \lstinline|local_list| and \lstinline|prio_queue|
and \lstinline|remaining_k| is decremented. If \lstinline|remaining_k| reaches zero, then the local
item list is appended to the global list, and all not yet seen (by this thread) items of the global
queue are added to the local priority queue.

\lstinline|DeleteMin| simply pops the local priority queue as long is it is non-empty and the popped
item has already been deleted. When a non-deleted item is popped, it is atomically marked as deleted
and returned to the caller. If instead we are faced with an empty local queue, we attempt to spy,
i.e. copy items from another thread's local list.

Both \lstinline|Insert| and \lstinline|DeleteMin| periodically synchronize with the global list
by adding all items that have not yet been seen locally to \lstinline|prio_queue|.
Memory allocations are handled using the wait-free memory manager by \citeauthor{wimmer2013wait} \cite{wimmer2013wait}.

The \citeauthor{wimmer2013data} priority queue was evaluated using a label-correcting variant
of Dijkstra's shortest path algorithms. Their model creates a task for each node
expansion, and therefore comes with a considerable task scheduling overhead. Nonetheless,
the parallel implementation scales well up to 10 threads, and further limited performance
gains are made until 40 threads. To date, no direct comparisons to other concurrent
priority queues have been possible since the data structure is strongly tied to \emph{Pheet},
but a separate implementation of k-priority queues is planned.

\subsection{SprayList} \label{sec:spraylist}

The SprayList is another recent approach towards a relaxed priority queue by \citeauthor{alistarhspraylist}
\cite{alistarhspraylist}. Instead of the distributed approach described in the previous section,
the SprayList is based on a central data structure, and uses a random walk in \lstinline|DeleteMin|
in order to spread accesses over the $O(P \log^3 P)$ first elements with high probability, where $P$
is again the number of participating threads.

\citeauthor{fraser2004practical}'s lock-free SkipList \cite{fraser2004practical} again serves as the
basis for the priority queue implementation. The full source code is available online\footnote{
\url{https://github.com/jkopinsky/SprayList}}. In the SprayList, \lstinline|Insert| calls are simply
forwarded to the underlying SkipList.

The \lstinline|DeleteMin| operation however executes a random walk, also called a \emph{spray}, the
purpose of which is to spread accesses over a certain section of the SkipList uniformly such that
collisions between multiple concurrent \lstinline|DeleteMin| calls are unlikely. This is achieved by
starting at some initial height, walking a randomized number of steps, descending a randomized number of levels,
and repeating this procedure until a node $n$ is reached on the lowest level. If $n$ is not deleted,
it is logically deleted and returned. Otherwise, a \emph{spray} is either reattempted, or the thread
becomes a cleaner, traversing the lowest level of the SkipList and physically removing logically deleted
nodes it comes across. A number of dummy nodes are added to the beginning of the list in order to counteract
the algorithm's bias against initial items.

The \emph{spray} parameters are chosen such that with high probability, one of the $O(P \log^3 P)$
first elements is returned, and that each of these elements is chosen roughly uniformly at random.
The final effect is that accesses to the data structure are spread out, reducing contention and resulting
in a noticeably lower number of \ac{CAS} failures in comparison to strict priority queues described
in Section \ref{sec:lockfree}.

The authors do not provide any statement as to the linearizability (or other concurrent correctness
criteria) of the SprayList, and it is not completely clear how to define it since no
sequential semantics are given.

Benchmarks show promising results: the SprayList scales well at least up to 80 threads,
and performs close (within a constant factor) to an implementation using a random remove instead
of \lstinline|DeleteMin|, which the authors consider as the performance ideal.


\section{Performance Results}

In this section, we compare the performance of several different concurrent priority
queue implementations:

\begin{itemize}
\item \textit{GlobalLock} An instance of the \lstinline|std::priority_queue<T>| class provided
      by the C++ standard library, protected by a single global lock.
\item \textit{Heap} \citeauthor{hunt1996efficient} provide an implementation
      of their Heap-based design \cite{hunt1996efficient} at \url{ftp://ftp.cs.rochester.edu/pub/packages/concurrent_heap/}.
      However, as the original code was written for the MIPS architecture, we chose to
      use the alternative implementation in \emph{libcds}\footnote{\url{http://sourceforge.net/p/libcds/code/}}
      instead. The heap capacity was $2^{18}$ --- lower values led to a quasi-deadlock situation under high concurrency
      as described in \cite{dragicevic2008survey}.
\item \textit{Noble} An implementation of the \citeauthor{sundell2003fast} priority queue \cite{sundell2003fast}
      is available in the Noble library\footnote{\url{http://www.non-blocking.com/Eng/download-demo.aspx}}.
      \textit{Noble} is lock-free, based on SkipLists, and limited to unique priorities.
      According to the authors, the commercial Noble library includes a more efficient
      version of this data structure that can also handle duplicate priority values.
\item \textit{Linden} Code for the \citeauthor{linden2013skiplist} priority queue \cite{linden2013skiplist}
      is provided by the authors under an open source license\footnote{\url{http://user.it.uu.se/~jonli208/priorityqueue}}.
      It is lock-free and uses \citeauthor{fraser2004practical}'s lock-free
      SkipList. The aim of this implementation is to minimize contention in
      calls to \lstinline|DeleteMin|. We chose $32$ as the \lstinline|BoundOffset| in order to optimize
      performance on a single socket. A \lstinline|BoundOffset| of $128$ performed only marginally better
      at high thread counts.
\item \textit{SprayList} A relaxed concurrent priority queue based on \citeauthor{fraser2004practical}'s
      SkipList using random walks to spread data accesses
      incurred by \lstinline|DeleteMin| calls. Code provided by \citeauthor{alistarhspraylist} is
      available on Github\footnote{\url{https://github.com/jkopinsky/SprayList}}.
\end{itemize}

Unfortunately, we were not able to obtain an implementation of the \citeauthor{shavit2000skiplist}
priority queue for benchmarking. The \citeauthor{wimmer2013data} data structure was omitted in the
following benchmarks since it is tightly coupled with the task-scheduling framework \emph{Pheet},
and cannot directly be compared to other implementations in its current form.

In each test run, the examined priority queue was initially filled with $2^{15}$
elements. We then ran a tight loop of $50\%$ insertions and $50\%$ deletions
for a total of $10$ seconds, where all \lstinline|Insert|
operations within this context implicitly choose a key uniformly at random from
the range of all 32-bit integers. This methodology seems to be the de facto standard for concurrent
priority queue benchmarks
\cite{alistarhspraylist,linden2013skiplist,shavit2000skiplist,sundell2003fast}.
Each run was repeated for a total of $10$ times
and we report on the average throughput.

All benchmarks were compiled with \verb|-O3| using GCC 4.8.2, and executed with threads pinned to
cores. Evaluations took place
on an 80-core Intel Xeon E7-8850 machine with each processor clocked at 2 GHz.
The benchmarking code was adapted from \citeauthor{linden2013skiplist}'s
benchmarking suite and is available at \url{https://github.com/schuay/seminar_in_algorithms}.

\begin{center}
\includegraphics[width = \textwidth]{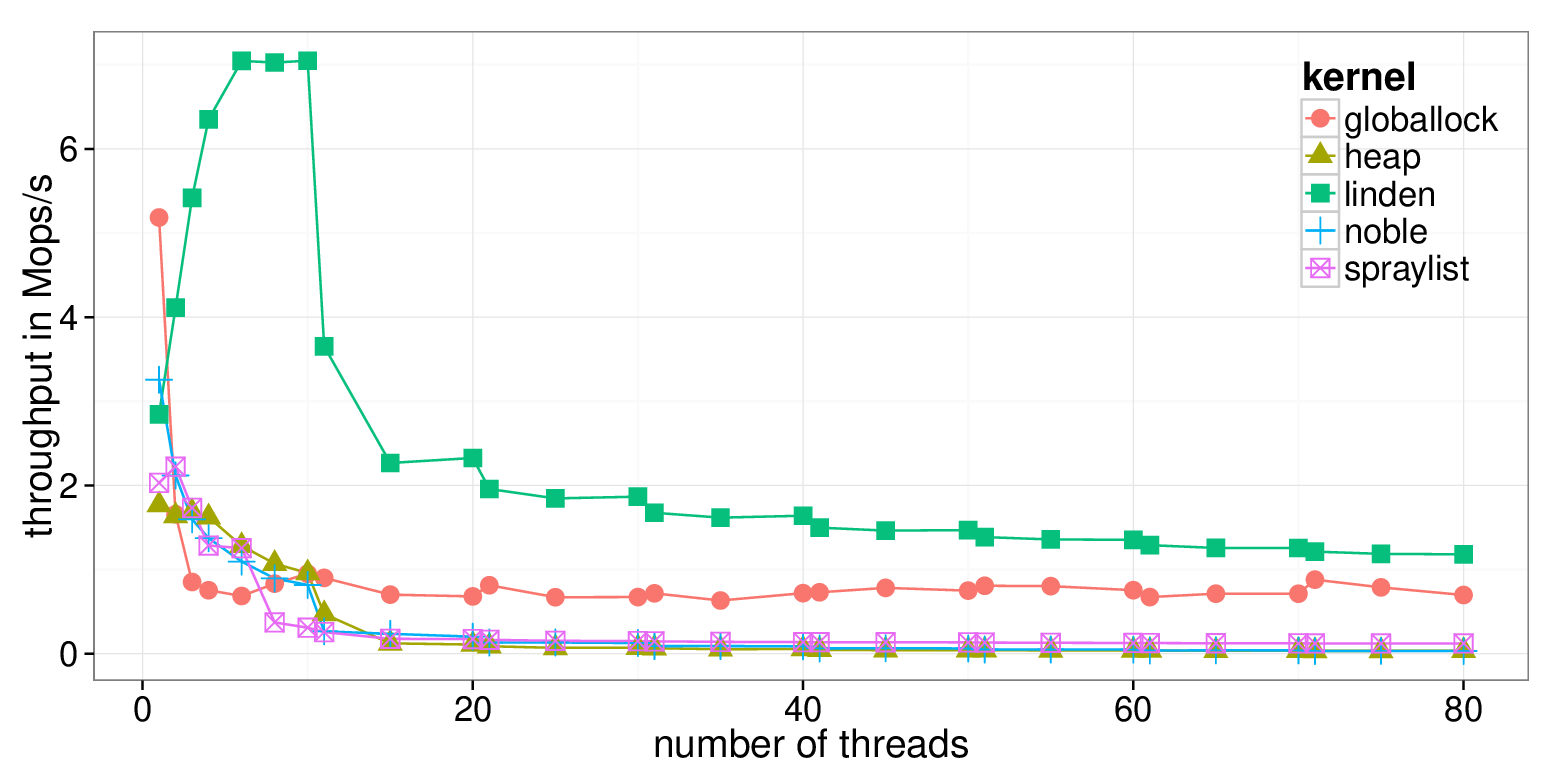}\end{center}

The \textit{Linden} priority queue emerged as the clear winner of these benchmarks,
with impressive scalability while all threads remain on the same socket. In fact, it is the only
tested implementation to scale to any significant degree; \textit{Heap} achieves minor gains
for up to 3 threads, while all others immediately lose throughput under concurrency.
Performance of the \textit{Linden} queue drops significantly once it is executed on more than 10
threads, thus incurring communication overhead across sockets. This effect is repeated to a lesser
degree every time an additional socket becomes active. However, the \textit{Linden} queue has
clearly superior throughput until the tested maximum of 80 threads.

\textit{GlobalLock} also performs surprisingly well at high concurrency levels.
It remains competitive when compared to \textit{Linden} at about $\frac{2}{3}$ of its throughput.
For single-threaded execution, it achieves the highest throughput of all tested implementations.
It is interesting to note that the \textit{GlobalLock} shows complementary behavior to
\textit{Linden} when new sockets become active; we believe that the additional latency incurred
by cross-socket communications might lead to reduced contention at the single lock.

All remaining implementations behave similarly, displaying very low throughput under concurrency.
This is a surprise in the case of \textit{Noble}; the original benchmarks performed by
\citeauthor{sundell2003fast} led us to expect improvements upon both \textit{GlobalLock}
and \textit{Heap} \cite{sundell2003fast}. However, we are somewhat reassured in our benchmarks by
the fact that \citeauthor{linden2013skiplist}'s results are similar.

The \textit{SprayList} result is even more unexpected, and we believe that it may be caused by
our use of \lstinline|malloc| instead of the included custom lock-free allocator, which we were
unable to execute without errors.

As a final observation we note that even the most efficient strict concurrent priority queue
never exceeds the throughput of the simple sequential heap executed on a single thread by more
than a factor of two. There is hope however since relaxed data structures continue to scale
beyond 10 threads (see the benchmarks performed in \cite{alistarhspraylist} for an example).

\section{Related Work} \label{sec:related}

There have been many publications on more specialized variants of concurrent priority queues
which were not covered in this paper.


\citeauthor{liu2012lock} present a lock-free, array-based priority queue in \cite{liu2012lock}
using the non-standard \ac{DCAS} and \ac{DCSS} instructions. In addition to the
\lstinline|Insert| and \lstinline|DeleteMin| operations, it also provides an \lstinline|ExtractMany|
function which returns the $n$ highest priority items in the queue, and a function that returns
items which ``probably'' have high priority. Other implementations based on non-standard instructions
can be found in \cite{israeli1993efficient,greenwald1999non}.


Another possibility is the restriction of priorities to a particular set, i.e. bounded range priorities.
\citeauthor{shavit1999scalable} present two such data structures in \cite{shavit1999scalable} using
combining funnels and bins of items. Experimental results have shown strong scalability until at least
256 threads.


Concurrent priority queues have also been studied in other contexts such as distributed memory systems.
\citeauthor{karp1993randomized} describe a distributed priority queue in which each processor owns
a local priority queue, inserts are sent to a random priority queue, and \lstinline|DeleteMin| operations
simply access the local queue \cite{karp1993randomized}.
\citeauthor{sanders1998randomized} extends this idea to remove the globally best elements instead
\cite{sanders1998randomized}.


Finally, the principle of relaxation has also been applied to other data structures.
\citeauthor{kirsch2012fast} invented an efficient k-FIFO queue \cite{kirsch2012fast},
The trend towards relaxed data structures in general is examined further in
\cite{shavit2011data,kirsch2012incorrect}.

\section{Conclusion} \label{sec:conclusion}

Priority queues are one of the most important abstract data structures in computer science,
and much effort has been put into parallelizing them efficiently. In this paper,
we have outlined the evolution of concurrent priority queues from initial heap-based designs,
through a period of increasingly efficient SkipList queues, to current research into relaxed data structures.

The switch from heaps to SkipLists as the backing data structure highlights how a simple change in
direction can help revitalize an entire field of research. SkipList-based priority queues are the current
state of the art in strict shared-memory concurrent priority queues: they provide strong guarantees
and scale well to up to the tens of threads in practice. Important limiting factors are contention
at the front of the list and the large number of \ac{CAS} failures.
The \citeauthor{linden2013skiplist} queue is designed
to minimize the latter; but the former is inherent to all strict priority queues, which could mean
that the peak performance in such structures has been reached.

Recently invented relaxed priority queues do not exhibit the inherent bottleneck at the front of the list,
as they do not necessarily return the minimal element within the queue and are able to spread
\lstinline|DeleteMin| accesses over a larger area of the structure. In consequence, relaxed queues scale noticeably
better and to larger thread counts than strict designs. Further research is necessary in order to fully
explore the possibilities provided by relaxed data structures.

\printbibliography

\end{document}

%% file: acronyms.tex
\acrodef{CAS}{Compare-And-Swap}
\acrodef{DCAS}{Double-Compare-And-Swap}
\acrodef{DCSS}{Double-Compare-Single-Swap}
\acrodef{DS}{Data Structure}
\acrodef{FAA}{Fetch-And-Add}
\acrodef{FAO}{Fetch-And-Or}
\acrodef{PQ}{Priority Queue}
\acrodef{SL}{SkipList}
\acrodef{TAS}{Test-And-Set}